\documentclass[useAMS,usedcolumn]{mn2e}
\usepackage{supertabular,epsf,rotate}
\newcommand{\f}{\phantom{2}}
\newcommand{\mc}{\multicolumn}

\newcommand{\ltsimeq}{\raisebox{-0.6ex}{$\,\stackrel
        {\raisebox{-.2ex}{$\textstyle <$}}{\sim}\,$}}
\newcommand{\gtsimeq}{\raisebox{-0.6ex}{$\,\stackrel
        {\raisebox{-.2ex}{$\textstyle >$}}{\sim}\,$}}

\input epsf.sty

\begin{document}

\title[A submillimetre difference between radio galaxies and radio quasars]
{A submillimetre difference between radio galaxies and radio quasars: evidence
for quasar-heated dust and synchronized submillimetre and radio source
activity}

\author[Willott et al.]{Chris J.\ Willott$^{1}$\footnotemark, Steve
Rawlings$^{1}$, Elese N. Archibald$^{2}$ \& James S. Dunlop$^{3}$ \\
$^{1}$ Astrophysics, Department of Physics, Keble Road, Oxford, OX1
3RH, U.K. \\ $^{2}$ Joint Astronomy Centre, 660 N. A`oh\={o}k\={u}
Place, University Park, Hilo, Hawaii, 9620, USA \\ $^{3}$ Institute
for Astronomy, University of Edinburgh, Blackford Hill, Edinburgh, EH9
3HJ, UK\\}

\maketitle

\begin{abstract}

We present submillimetre photometry of 11 3CR and 6CE radio quasars of
similar 151-MHz radio luminosity and redshifts to the radio galaxies
studied at $1.3<z<2$ by Archibald et al.\ (2001).  We detect all bar
one quasar at 850 $\umu \rm m$, and five quasars are confirmed as
dusty hyperluminous objects by detections at shorter wavelengths.  Our
observations reveal a clear difference between the submillimetre
luminosity distributions of the radio quasars and a matched sample of
radio galaxies: the quasars are $\sim 5$-times brighter than the radio
galaxies and $\geq 2$-times brighter accounting for possible
synchrotron contamination. This difference is in quantitative
agreement with a receding-torus unified scheme for radio sources in
which the torus opening angle depends on quasar optical luminosity,
provided that there is a close relationship between the optical and
submillimetre luminosities. The implication is that quasar-heated dust
dominates the submillimetre emission for powerful radio quasars at $z
\sim 1.5$.  We find a significant anti-correlation between
submillimetre/far-infrared luminosity $L_{\rm FIR}$ and radio source
age in the sense that hyperluminous quasars tend to be associated with
young ($<10^7$ yr) sources. This suggests that the processes
controlling $L_{\rm FIR}$ are synchronized with the jet-triggering
event and that $L_{\rm FIR}$ declines on a $\sim 10^{7}$ yr
timescale. There is evidence for a small amount of obscuration in the
hyperluminous quasars from reddening and associated or broad
absorption lines. We speculate that shortly after jet triggering, dust
is more widely distributed than at later times and is heated by the
quasar nucleus and/or a short-lived synchronized starburst. Any such
starburst would form only a few per cent of the total stellar mass in
agreement with the evidence for dominant old stellar populations in $z
\sim 1.5$ radio galaxies. In contrast, high-redshift ($z>3$) radio
galaxies with similar submillimetre luminosities could have longer
star-formation timescales due to the greater availability of gas and
could be forming the bulk of their stellar populations.

\end{abstract}


\footnotetext{Email: cjw@astro.ox.ac.uk}

\section{Introduction}

Submillimetre observations provide an excellent method of detecting
dusty star-forming galaxies at high redshifts: UV radiation from hot
stars is absorbed by dust, re-radiated in the far-infrared and
redshifted into the submillimetre. The steep submillimetre spectral
index leads to large $k$-corrections which, at high redshifts,
compensates for increasing cosmological dimming so that an object with
a given submillimetre luminosity has a similar observed flux-density
at $z=1$ to a similarly-luminous one at $z=5$ (Blain \& Longair 1993).

Powerful radio sources can be easily identified out to high redshifts
and are thought to be associated with massive elliptical galaxies, or
their progenitors. The existence of this close association is well
known at low-to-intermediate redshifts (e.g.\ Owen \& Laing 1989) and
it is likely to be maintained to higher redshifts by scaling relations
between radio luminosity, accretion luminosity, black hole mass and
(dark matter) halo mass (e.g.\ Rawlings \& Saunders 1991; Magorrian et
al. 1998).  At low-redshifts, radio galaxies appear to be
kinematically-relaxed massive ellipticals (McLure et al. 1999; Dunlop
et al. 2001; Bettoni et al. 2001) with old stellar populations (Nolan
et al. 2001). These facts, together with the small scatter in the
near-infrared Hubble diagram to $z\approx 3$ (Jarvis et al. 2001a),
suggest that most of the star formation in powerful radio galaxies
occurred at early cosmic epochs.

In the first large study of the submillimetre emission of
high-redshift radio galaxies, Archibald et al. (2001; hereafter A01)
made sensitive (850 $\umu$m sensitivity $\approx 1$ mJy rms)
submillimetre observations of 47 radio galaxies with the SCUBA array
on the JCMT. Their sample covered the redshift range $1<z<4.5$ and
they detected dust emission in 14 objects. The main result of their
study was a clear difference between the detection rate at $z<2.5$ of
15\% and the detection rate at $z>2.5$ of 75\%. This indicates that
the submillimetre luminosity of radio galaxies increases
systematically with redshift, suggesting that the bulk of star
formation in radio galaxies, and by inference in all massive galaxies,
occurred at epochs corresponding to high redshifts.

\begin{table*}
\begin{center}
%\tiny
\begin{tabular}{lcccrcrrrcc}
\hline\hline
%\vspace{0.2cm}
 \mc{1}{l}{ Name} & \mc{1}{c}{Class}& \mc{1}{c}{Redshift}& \mc{1}{c}{$\log L_{151}$}& \mc{1}{c}{$S_{\rm core}$}& \mc{1}{c}{$\log L_{\rm [O\,{\small III}]}$}& \mc{1}{c}{$D$}& \mc{1}{c}{ $S_{850}$}   &\mc{1}{c}{ $S_{450}$}  &\mc{1}{c}{ $S_{850}$}  & \mc{1}{c}{$S_{450}$}\\
 \mc{1}{l}{ } & \mc{1}{c}{ }& \mc{1}{c}{$z$}& \mc{1}{c}{W/Hz/sr}& \mc{1}{c}{mJy}& \mc{1}{c}{W}& \mc{1}{c}{ kpc }& \mc{1}{c}{(mJy)}  & \mc{1}{c}{(mJy)} & \mc{1}{c}{$2\sigma$ limit}  & \mc{1}{c}{$2\sigma$ limit}\\
\hline\hline

3C 181       &  Q &  1.382 &  28.11 &   6.0  &  37.37 &  48.0 & $ {\bf 5.27}\pm 1.06$ & $      -0.8 \pm \f 7.5 $ & --   & \f 15 \\
3C 191       &  Q &  1.952 &  28.52 &  38.3  &  37.21 &  41.1 & $ {\bf 6.39}\pm 1.06$ & ${\bf 28.9} \pm \f 8.1 $ & --   & \f -- \\
3C 205       &  Q &  1.534 &  28.19 &  24.6  &  37.00 & 152.4 & $ {\bf 2.36}\pm 1.10$ & $      11.6 \pm \f 7.9 $ & --   & \f 27 \\
3C 268.4     &  Q &  1.400 &  27.97 &  42.9  &  37.25 &  91.9 & $ {\bf 5.13}\pm 1.18$ & $     9.9   \pm   12.5 $ & --   & \f 35 \\
3C 270.1     &  Q &  1.519 &  28.24 &  60.0  &  36.51 & 101.6 & $ {\bf 7.42}\pm 1.17$ & $      15.4 \pm   11.5 $ & --   & \f 38 \\
3C 280.1     &  Q &  1.659 &  28.36 &  36.0  &  36.72 &  39.8 & $ {\bf 5.10}\pm 1.74$ & $       0.4 \pm   23.8 $ & --   & \f 48 \\
3C 298       &  Q &  1.439 &  28.57 &  515   &  37.26 &  12.7 & ${\bf 21.13}\pm 2.06$ & $      -8.0 \pm   23.7 $ & --   & \f 47 \\
3C 318       &  Q &  1.574 &  28.11 &  22.0  &  37.18 &   6.8 & $ {\bf 7.78}\pm 1.00$ & ${\bf 21.6} \pm   10.6 $ & --   & \f -- \\
3C 432       &  Q &  1.785 &  28.31 &   7.5  &  36.68 & 109.8 & ${\bf 7.93} \pm 1.70$ & $      2.91 \pm   20.9 $ & --   & \f 45 \\
6C 0955+3844 &  Q &  1.405 &  27.45 &   4.3  &  36.58 & 181.3 & $    -0.01  \pm 1.11$ & $       8.9 \pm \f 7.8 $ & 2.22 & \f 25 \\
6C 1045+3513 &  Q &  1.594 &  27.22 &  10.0  &  35.94 &   1.7 & ${\bf 9.18} \pm 1.10$ & ${\bf 21.3} \pm \f 6.8 $ & --   & \f -- \\
4C 13.66     &  G &  1.450 &  28.09 &   0.8  &  36.63 &  50.7 & ${\bf 3.53} \pm 0.96$ & $   -16.2   \pm   18.2 $ & --   & \f 36 \\
3C 239       &  G &  1.781 &  28.42 &   0.5  &  36.38 &  94.6 & $    0.83   \pm 1.00$ & $    -2.1   \pm   18.3 $ & 2.83 & \f 37 \\
3C 241       &  G &  1.617 &  28.11 &   5.0  &  36.75 &   7.7 & $    1.81   \pm 0.94$ & $    14.8   \pm   12.6 $ & 3.69 & \f 40 \\
3C 294       &  G &  1.786 &  28.39 &   0.5  &  36.97 & 126.7 & $    0.19   \pm 0.78$ & $    5.4    \pm   13.4 $ & 1.75 & \f 32 \\
3C 322       &  G &  1.681 &  28.21 &   2.2  &  35.97 & 279.4 & $   -0.05   \pm 1.06$ & $   -37.5   \pm   16.0 $ & 2.12 & \f 32 \\
3C 437       &  G &  1.480 &  28.11 &   0.1  &  36.38 & 291.0 & $   -1.18   \pm 0.98$ & $    2.9    \pm   17.3 $ & 1.96 & \f 37 \\
3C 470       &  G &  1.653 &  28.15 &   2.0  &  35.93 & 203.2 & ${\bf 5.64} \pm 1.08$ & $    57.8   \pm   32.9 $ & --   &   124 \\
6C 0820+3642 &  G &  1.860 &  27.55 &   1.0  &  36.00 & 193.8 & ${\bf 2.07} \pm 0.96$ & $    13.6   \pm   18.0 $ & --   & \f 50 \\
6C 0901+3551 &  G &  1.904 &  27.46 &   0.5  &  36.12 &  21.9 & $   -1.83   \pm 1.15$ & $    -19.3  \pm \f 8.2 $ & 2.30 & \f 16 \\
6C 0905+3955 &  G &  1.882 &  27.71 &   1.1  &  35.76 & 934.5 & ${\bf 3.62} \pm 0.89$ & $    31.2   \pm   16.2 $ & --   & \f 64 \\
6C 0919+3806 &  G &  1.650 &  27.53 &   6.0  &  34.96 &  88.1 & $   -0.88   \pm 1.05$ & $    10.5   \pm   10.1 $ & 2.10 & \f 31 \\
6C 1204+3708 &  G &  1.779 &  27.61 &   1.0  &  36.45 & 435.2 & $    0.16   \pm 1.25$ & $    45.1   \pm   26.6 $ & 2.66 & \f 98 \\

\hline\hline
\end{tabular}
\end{center}

{\caption[Table]{\label{tab:obs} Submillimetre observations and basic
data for the sample of radio quasars (class `Q') and radio galaxies
(`G') in the redshift range $1.37 < z< 2.0$ described in Sec.\
2. Detections at the $\geq 2\sigma$ level are given in bold
type. Submillimetre flux-densities of the radio galaxies are from
A01. No corrections have been made for possible contributions of
synchrotron emission to the submillimetre fluxes. Radio core fluxes
are quoted at 5 GHz (observed-frame) in most cases. Radio data for the
radio galaxies and quasars are from the references listed in A01 and
Blundell et al. (in prep.) respectively. Data for the [O\,{\small
III}] emission line luminosities have been taken from Jackson \&
Rawlings (1997) for 3CR sources and Rawlings et al. (2001) for 6C
sources, apart from these exceptions: 3C~241 -- Hirst, Jackson \&
Rawlings (2002); 3C~270.1 and 6C~0955+3844 -- Willott et al. (in
prep.); 3C~298 -- Espey et al. (1989); 3C~318 -- Willott et
al. (2000b); 6C~1045+3513 (this paper).  For some sources, [O\,{\small
III}] has not been observed and a different narrow line flux is used
and the [O\,{\small III}] flux estimated from the line ratios in
McCarthy (1993). These sources are 3C~298 ([O\,{\small II}]);
6C~0820+3642 (Ly$\alpha$); 6C~0901+3551(Ly$\alpha$); 6C~0905+3955
(Ly$\alpha$); 6C~0919+3806 (C\,{\small II}]); 6C~1045+3513
([Ne\,{\small V}]); 6C~1204+3708 (Ly$\alpha$). For the quasar 3C~432,
no narrow line data have been published and the [O\,III] flux has been
estimated assuming a rest-frame equivalent width of 30 \AA~
(e.g. Miller et al. 1992).  }} \end{table*}

Unified schemes for radio galaxies and radio quasars\footnotemark
propose they are the same objects viewed along different
lines-of-sight, such that in radio galaxies the nuclear regions are
obscured by a dusty torus or warped disc (Scheuer 1987; Barthel
1989). Such theories have proved remarkably successful (see Antonucci
1993 for a review), at least at high radio luminosities (Willott et
al. 2000a). For objects at redshifts $z\approx 1.5$, the observed 850
$\umu$m radiation was emitted at a rest-frame wavelength of $\approx
300~\umu$m. At these wavelengths, dust is expected to be optically
thin and its thermal emission isotropic (e.g. Granato \& Danese 1994;
Efstathiou \& Rowan-Robinson 1995). Therefore, according to the
simplest type of radio-loud AGN unification scheme, the submillimetre
properties of radio galaxies and radio quasars, matched in terms of
low-frequency radio luminosity and redshift, should be similar. In
more realistic unified schemes, however, they may not be expected to
be identical.  Rawlings \& Saunders (1991) and Simpson (1998) have
argued that samples of radio galaxies and radio quasars, matched in
one orientation-independent quantity (i.e.\ low-frequency radio
luminosity), are expected to have different distributions in a second
orientation-independent quantity (i.e.\ [O\,{\small III}] emission
line luminosity) if the two quantities are linked by a scaling
relation with intrinsic scatter.  The first motivation behind the work
described in this paper was to explore unified schemes further by
looking for significant differences between the submillimetre
properties of radio galaxies and radio quasars matched in 151-MHz
luminosity and redshift.

\footnotetext{ We will use the term `radio quasars' to refer to
lobe-dominated steep-spectrum radio sources associated with quasar
nuclei. These are the dominant population of radio-loud quasars in
low-frequency-selected samples such as 3CRR and 6CE, whereas compact
flat-spectrum quasars dominate in high-frequency-selected
samples. Flat-spectrum quasars often have high submillimetre fluxes
simply by virtue of a continuation of their synchrotron spectra from
radio to shorter wavelengths.}

The second motivation behind this paper was to explore further the
suggestion of Blundell \& Rawlings (1999) that some part of the
evolution in the submillimetre properties of radio galaxies with
redshift is a selection effect linked to the inevitable youth of the
highest-redshift radio sources from flux-limited samples.  A01 have
already considered this suggestion, arguing that it is not a primary
driver of the observed redshift dependencies.  However, with a small
dynamic range in source age, and somewhat uncertain selection effects,
the high-redshift sample of Archibald et al.\ is not ideally suited to
a test of the idea that submillimetre properties might be correlated
with the time since the jet-triggering event.

At $z \sim 1.5$ A01 have already obtained submillimetre data on an
unbiased sample of 3CRR and 6CE galaxies, and we aimed to complement
these with new observations of a matched sample of radio quasars.  In
Section 2 we describe the observations and results. In Section 3 we
consider the likely contribution to the measured submillimetre fluxes
from an extension of the radio synchrotron spectrum. In Section 4 we
consider explanations for our results. We present our conclusions in
Section 5. The convention for all spectral indices $\alpha$ is that
flux-density $S_{\nu} \propto \nu^{-\alpha}$. We assume throughout
that $H_{0}=70~ {\rm km~s^{-1}Mpc^{-1}}$, $\Omega_{\rm M}=0.3$ and
$\Omega_ {\Lambda}=0.7$.

\section{Observations}

We have defined a sample of radio quasars in the redshift range
$1.37<z<2.0$ selected from two complete radio samples (3CRR -- Laing,
Riley \& Longair 1983; 6CE -- Rawlings, Eales \& Lacy 2001) to compare
with the radio galaxies from the same samples in the same redshift
range in the study of A01. Due to scheduling reasons only 3CRR quasars
within the Right Ascension range $04^{\mathrm h} < {\mathrm {RA}} <
22^{\mathrm h}$ are included in our sample. These were supplemented
with two quasars from the revised 3C catalogue (3CR -- Spinrad et
al. 1985). These two quasars, 3C~298 and 3C~280.1, are not included in
the 3CRR sample due to reasons of low declination and radio
flux-density just below the 3CRR limit, respectively. They are the
only high galactic latitude quasars in the target redshift range in
the 3CR sample which are not included in the 3CRR sample. The total
sample consists of 11 quasars and the comparison sample extracted from
A01 contains 12 radio galaxies (Table 1).

The sample of radio quasars was observed in photometry mode with the
SCUBA bolometer array at the JCMT in March, April, October 1999,
November 2000, February and April 2001. Observations were made
simultaneously at 850 $\umu$m and 450 $\umu$m and the sensitivities
reached at 850 $\umu$m are $\sim 1$ mJy rms. The atmospheric zenith
opacity at 850 $\umu$m ranged from 0.16 to 0.52 with a median of
0.22. The data were reduced using the SURF package, following the same
method as in A01 for consistency. The resulting flux-densities are
given in Table 1 along with the data for the matched radio galaxy
sample. Ten out of eleven radio quasars were detected at 850 $\umu$m
at $> 2 \sigma$ significance (eight at $> 3 \sigma$), compared with
only four $> 2 \sigma$ detections out of the twelve radio galaxies. At
the shorter wavelength of 450 $\umu$m, three quasars have $> 2 \sigma$
detections, compared to none of the radio galaxies (although the
quasar observations are generally more sensitive at 450 $\umu$m than
those of the radio galaxies due to the introduction of new filters in
October 1999). The mean quasar $S_{850}$ is $6.3 \pm 1.0$ mJy and the
mean radio galaxy $S_{850}$ is $1.2 \pm 0.6$ mJy. Therefore the
quasars are brighter than the radio galaxies in the submillimetre by a
factor of 5.

Relating the observed submillimetre flux-density to a
submillimetre/far-infrared luminosity is not a trivial matter, given
the poor constraints on the shape of the dust spectra for most sources
which have only been detected at one or two wavelengths. Therefore, we
adopt a template dust spectrum as has been determined from
observations of a sample of similarly luminous
(submillimetre/far-infrared luminosities $L_{\rm FIR} \sim
10^{13}L_{\sun}$) objects by Priddey \& McMahon (2001). This template
consists of isothermal optically-thin dust with temperature $T=41$ K
and emissivity index $\beta=1.95$.  For our adopted cosmological model
and dust template, the observed $850\umu$m flux-density $S_{850}$ of a
source with a given $L_{\rm FIR}$ does not change appreciably over the
redshift range $z=1$ to $z=5$. The conversion between $L_{\rm FIR}$
and $S_{850}$ for our adopted dust spectrum is $L_{\rm FIR} \approx 2
\times 10^{12} (S_{850} / {\rm mJy}) L_{\sun}$. However, for higher
dust temperatures or multi-temperature models (which are required to
fit the infra-red SEDs of most luminous AGN -- e.g. Rowan-Robinson
2000), $L_{\rm FIR}$ could be up to an order of magnitude greater than
this, so these values of $L_{\rm FIR}$ should be treated as lower
limits. Therefore the radio quasars we have detected with SCUBA have
lower limits on their far-infrared luminosities in the range $4 \times
10^{12}$ to $2 \times 10^{13} L_{\sun}$. It is likely that many of
them would be classified as hyperluminous infrared sources ($L_{\rm
FIR} \geq 10^{13}L_{\sun}$) if their far- and mid-infrared fluxes were
measured. Throughout most of this paper we will use the observed
submillimetre flux-density in preference to the far-infrared
luminosity and thereby avoid uncertainty in $L_{\rm FIR}$ due to the
unknown dust spectral shape.

Five of the radio quasars are also detected at wavelengths shorter
than 850$\umu$m. 3C~191, 3C~318 and 6C~1045+3513 are detected at
450$\umu$m with SCUBA and 3C~268.4, 3C~298 and 3C~318 are detected in the
far-infrared by ISO and/or IRAS. These objects are discussed in further
detail here.

\subsection{Notes on objects detected at $100-450 \umu$m}

\vspace{0.2cm}
\noindent {\bf 3C~191}

The submillimetre spectral index of 3C~191 determined from the
850$\umu$m and 450$\umu$m fluxes is $-2.4 \pm 0.6$, consistent within $2
\sigma$ with the value of $-3.2$, expected for our assumed dust
template. We find that $L_{\rm FIR} > 1.2 \times 10^{13}L_{\sun}$. The
optical spectral slope is $\alpha_{\rm opt}=0.7$ (Barthel, Tytler \&
Thomson 1990), quite similar to the typical radio quasar value of
$\alpha_{\rm opt}=0.5$ (Brotherton et al. 2001). This indicates that
there is little reddening of the quasar light along our
line-of-sight. However, 3C~191 has very strong associated absorption
lines with a C\,{\small IV} absorption equivalent width of 6.1 \AA~(Anderson et
al. 1987). Hamann et al. (2001) studied these absorption systems with
high-resolution spectroscopy and determined the distance of the
absorbers from the nucleus to be $\approx 30$ kpc with a corresponding
characteristic flow time of $3 \times 10^{7}$ years. The distances of
the radio hotspots from the nucleus are of a similar size-scale
(projected linear size=41 kpc), corresponding to an approximate age of
the radio source of $3 \times 10^{6}$ years (see Sec.\ 4.2).

\vspace{0.2cm}
\noindent {\bf 3C~268.4}

\begin{figure}
\vspace{0.7cm}
\epsfxsize=0.48\textwidth
\epsfbox{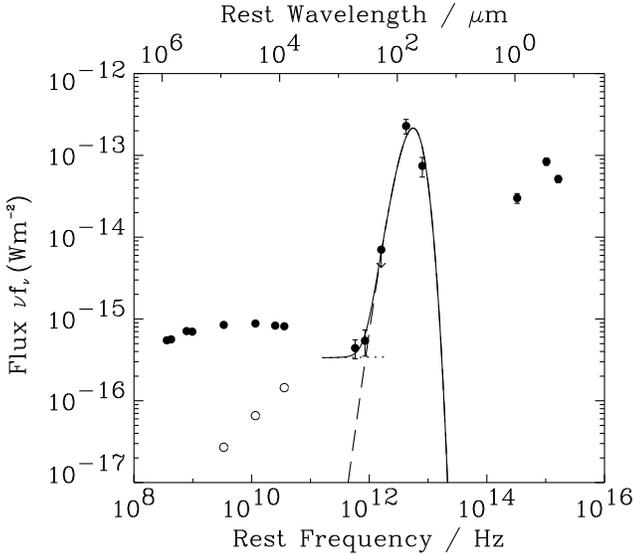}

{\caption[junk]{\label{fig:3c268p4} Rest-frame SED from radio to
optical of the hyperluminous IR quasar 3C~268.4. For observations
which separate the core emission from that of the extended radio
components, the core flux is shown as an open circle. The dotted line
is a $\alpha = 1$ core spectrum which dominates the millimetre
flux. The dashed curve is the best fit thermal dust spectrum ($T=48$
K, $\beta=2.5$, optically-thin at $\lambda > 50\umu$m). The solid
curve shows this added to the synchrotron core spectrum. }}

\end{figure}

3C~268.4 is detected at $850 \umu$m ($5.1 \pm 1.2$ mJy) but not at
$450 \umu$m. Recent detections at $170 \umu$m and $90 \umu$m with ISO
and at 1.3 mm with IRAM ($6.1 \pm 1.6$ mJy) indicate a high far-IR
luminosity and some synchrotron contamination of the mm fluxes
(Andreani et al. 2001). Estimating the millimetre contamination by
synchrotron emission is not a trivial matter due to the lack of other
data at frequencies higher than 15 GHz (Fig.\, \ref{fig:3c268p4}). For
reasons to be discussed in Sec.\ \ref{sync} we expect the extended
radio lobe spectrum to steepen sharply at high frequencies and the
main synchrotron component in the millimetre regime to be the compact
core. It is also clear in Fig.\, \ref{fig:3c268p4} that the core
spectrum also steepens between 15 GHz and 1.3 mm. To estimate the
synchrotron contamination to the $850 \umu$m flux-density we assume a
millimetre core spectrum with $\alpha=1$ which gives $S_{850} = 3$
mJy. As we will show below, the properties of the dust spectrum
require a contribution to the $850 \umu$m flux-density from dust which
is in approximate agreement with this contribution from the core.

We have fitted the far-IR and sub-mm data with a range of thermal dust
spectra. Using only the ISO fluxes and the $450 \umu$m upper limit,
the strongly peaked spectrum at $\approx 170 \umu$m (observed frame)
provides strong constraints on the dust temperature and emissivity
index (Fig.\, \ref{fig:3c268p4}). The best fitting values are $T=48$ K
and $\beta=2.5$ for dust which is optically-thin at $\lambda >
50\umu$m and this spectrum gives an expected $850 \umu$m flux-density
of 2.7 mJy. The far-IR luminosity of this dust spectrum is $2 \times
10^{13} {\rm L_{\sun}}$. This value for the emissivity index is larger
than usually observed (Priddey \& McMahon 2001; Dunne et al. 2000) but
requiring $\beta$ to be no larger than 2.0 gives a similar far-IR
luminosity and a higher expected $850 \umu$m flux-density of 4.0 mJy.
Therefore we conclude that the dust contribution to the $850 \umu$m
flux-density is significant and independent of its actual value,
$L_{\rm FIR} >2 \times 10^{13} {\rm L_{\sun}}$ and the object is
classified as hyperluminous. 3C~268.4 is not heavily reddened in the
rest-frame UV but does show strong C\,{\small IV} associated
absorption (equivalent width $>1.9$ \AA; Anderson et al. 1987).

\vspace{0.2cm}
\noindent {\bf 3C~298}

\begin{figure}
\vspace{0.7cm}
\epsfxsize=0.48\textwidth
\epsfbox{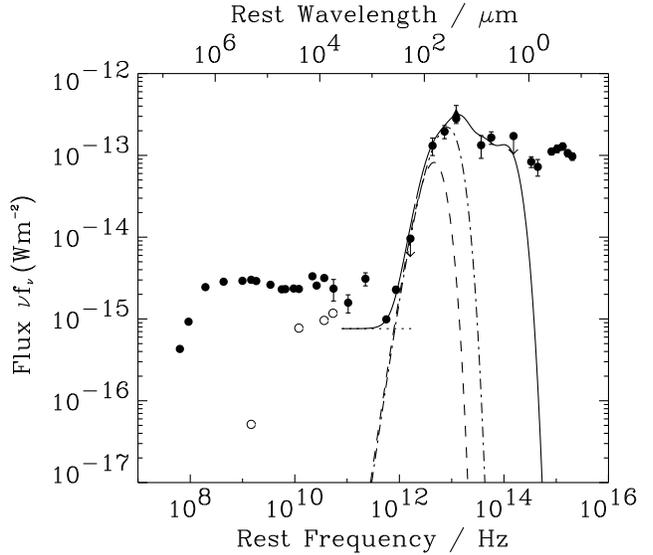}
{\caption[junk]{\label{fig:3c298} Rest-frame SED of the hyperluminous
IR quasar 3C~298 -- symbols as for Fig.\, \ref{fig:3c268p4}. The
dotted line is a $\alpha = 1$ core spectrum which dominates the
millimetre flux. The solid curve shows this added to the thermal IR
spectrum which is modelled as a combination of greybodies of
temperatures ranging from 50 K (this component is shown by a dashed
line) to 1300 K. The dot--dashed line shows that the
submillimetre/far-infrared data points could equally well be fitted by
a T=100 K dust spectrum.}}
\end{figure}

The source with the highest 850 $\umu$m flux-density in our sample is
3C~298. This quasar is extremely bright at all wavelengths from the
radio through to X-rays. It was detected by IRAS at 60 $\umu$m and by
ISO at five wavelengths ranging from 12.8 to 170 $\umu$m (Meisenheimer
et al. 2001). 3C~298 has a strong radio core (515 mJy at 5 GHz; Akujor
et al. 1991) but is lobe-dominated at GHz frequencies. The measured
1.3 mm flux-density of 14.1 mJy (Meisenheimer et al. 2001) shows that
the core is still strong at millimetre wavelengths and makes a
substantial contribution to the observed 850 $\umu$m flux-density. We
have fitted the SED of 3C~298 to attempt to determine the relative
contributions of dust and synchrotron emission (Figure
\ref{fig:3c298}). This procedure is not trivial because the core
spectrum appears to be fairly flat up to a rest-frame frequency of 200
GHz and then steepens sharply to $\alpha \approx 2$ at higher
frequencies. However, the flux-density measurements at these high
frequencies have large errors and are only $1 \sigma$ from being
consistent with $\alpha = 1$, as is more common above the break in the
core spectra of lobe-dominated quasars (e.g. Antonucci, Barvainis \&
Alloin 1990; van Bemmel \& Bertoldi 2001). Therefore we assume a core
with $\alpha = 1$ over the observed 1300 to 850 $\umu$m range. We
conclude that $\approx 8$ mJy of the 850 $\umu$m flux-density is due
to synchrotron radiation and $\approx 13$ mJy is due to thermal dust
emission. The reason why this quasar was not detected at 450 $\umu$m
was due to the poor sensitivity caused by the high airmass of the
observation. The submillimetre and far-infrared data can be equally
well fit by thermal spectra with temperatures $T=50$ K (typical of
starbursts) and $T=100$ K. The total IR luminosity of 3C~298 is $8
\times 10^{13} {\rm L_{\sun}}$.

3C~298 is a small radio source with projected linear size of 12.7 kpc.
It shows very strong CIV associated absorption with equivalent width
4.5 \AA~(Anderson et al. 1987). The optical spectral slope is
$\alpha_{\rm opt}=1.5$, which is redder than typical quasars. This
reddening is most likely due to dust associated with the absorbing gas.

\vspace{0.2cm}
\noindent {\bf 3C~318}

In Willott, Rawlings \& Jarvis (2000b) we have already studied the
spectral energy distribution (SED) of 3C~318 and determined a total IR
luminosity of $>7 \times 10^{13} {\rm L_{\sun}}$ (corrected for the
cosmology used in this paper). The quasar optical spectrum is found to
be reddened, indicating the presence of dust in the quasar
environment. 3C~318 is a small (6.8 kpc) radio source.

\vspace{0.2cm}
\noindent {\bf 6C~1045+3513}

6C~1045+3513 is detected at both 850 and 450 $\umu$m, with a
submillimetre spectral index of $-1.3 \pm 0.7$. On the basis of the 850
$\umu$m observations we find $L_{\rm FIR} > 2 \times
10^{13}L_{\sun}$. In the Appendix we present new optical spectroscopy
of this object, showing it to be a reddened quasar, similar to
3C~318. The spectrum also shows a CIV broad absorption
feature. 6C~1045+3513 is the smallest radio source of the entire
sample of quasars and radio galaxies with a projected linear size of
just 1.7 kpc.

\begin{figure}
%\hspace{-0.7cm}
\epsfxsize=0.45\textwidth
\epsfbox{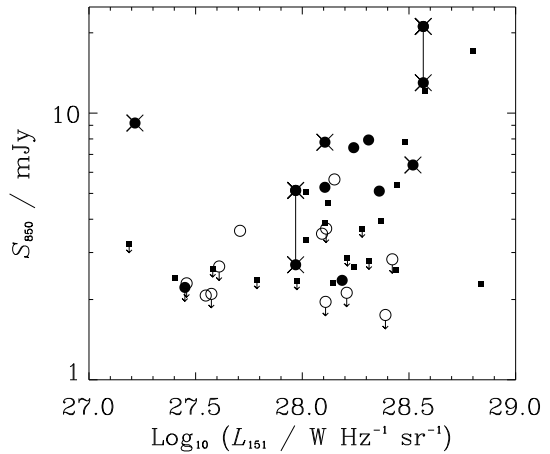}
{\caption[junk]{\label{fig:l151} 850 $\umu$m flux-density $S_{850}$
against low-frequency (151 MHz) radio luminosity $L_{151}$ for the
$1.37<z<2.0$ radio quasars (filled circles) and radio galaxies (open
circles). $2\sigma$ upper limits are denoted by symbols with
arrows. The five quasars with detections at shorter wavelengths
($100-450 \umu$m) have crosses through them. 3C~298 and 3C~268.4 are
shown as two symbols joined by a line, indicating the total
flux-density and that corrected for non-thermal emission. Also plotted
are radio galaxies at higher redshifts ($2<z<4.5$) from A01 (small
squares).}}
\end{figure}

\vspace{0.2cm}

\subsection{Comparison of radio quasars and radio galaxies}

Fig. \ref{fig:l151} plots the observed 850 $\umu$m flux-density
$S_{850}$ against low-frequency (151 MHz) radio luminosity $L_{151}$
for the matched quasar and radio galaxy samples. The main point to
note from Fig. \ref{fig:l151} is that there is a clear difference in
the flux-density distributions of the quasar and radio galaxy samples,
with the mean quasar submillimetre flux brighter by a factor of
5.\footnotemark

\footnotetext{A second point to note from Fig. \ref{fig:l151} is that
there is only marginal evidence for a correlation between $L_{151}$
and $S_{850}$. The survival analysis probability for a correlation
being present is $\approx 80$\%. Removing the outlying point of
6C~1045+3513 increases the correlation probability to $\approx
95$\%. This matter has been discussed by Archibald et al.\ (2001), and
we defer any further discussion to a wider investigation of the
dependences of submillimetre luminosity on radio luminosity and
redshift which will use new SCUBA data on radio sources from fainter
samples than the 3CR and 6CE samples used here (Rawlings et al., in
prep.).}

Due to the large number of 850 $\umu$m upper limits for the radio
galaxies, we use survival analysis statistical tests which can account
for these limits (Feigelson \& Nelson 1985; Isobe, Feigelson \& Nelson
1986). Using a variety of tests to compare the 850 $\umu$m flux
distributions of the quasars and radio galaxies (the Gehan, logrank
and Peto-Prentice tests), the flux-density distributions are found to
be different at $>99.9$\% significance.  Given the high fraction of
upper limits for the radio galaxies, we cannot unequivocally determine
the factor by which the quasars are brighter than the radio galaxies
in the submillimetre. Using the above tests we find that the quasars
are brighter than the radio galaxies by a factor $\geq 2$ at the 95\%
confidence level.

\section{The synchrotron contribution to submillimetre fluxes}
\label{sync}

We have found the robust result that radio quasars have $2-5$-times
higher submillimetre fluxes than radio galaxies of the same radio
luminosity and redshift. If this emission is due to optically-thin
dust, then this is inconsistent with the notion of radio quasars and
radio galaxies being identical apart from some orientation
dependence. Given that the radio sources we are discussing are some of
the most powerful known, it is essential that we consider whether the
observed submillimetre fluxes could be affected by an extrapolation of
the nonthermal radio synchrotron emission. Radio lobes and hotspots
generally have a very steep spectrum at millimetre wavelengths and are
unlikely to contribute significantly for many sources. This issue is
discussed in detail in A01 where they find that extended synchrotron
components very rarely contribute more than 1 mJy to the 850 $\umu$m
flux-density. However, A01 note that it is possible that two of the
radio galaxy detections in our sample (3C~470 and 6C~0820+3642) become
non-detections after correcting for synchrotron emission. We do not
attempt to make any corrections for this in this paper, due to the
large uncertainties involved in extrapolating the spectrum from radio
frequencies. In Sec.\ \ref{age} we will discuss this in terms of the
angular extent of the radio sources with respect to the size of the
SCUBA beam. Of course, radio quasars and radio galaxies would suffer
equally from lobe or hotspot emission, so it cannot explain the
difference in submillimetre properties between the two types of
object.

The emission from radio cores and jets is beamed and Doppler boosting
makes it highly orientation dependent. In the standard unification
scheme, radio quasars have their jet axes closer to our line-of-sight
and the enhanced Doppler boosting means that quasars have brighter
cores than radio galaxies. Note that because the 3C and 6C radio
samples were selected at low frequencies, where extended emission
dominates, the quasars in our sample are lobe-dominated and not like
the core-dominated blazars which are known to have fairly flat ($0
\leq \alpha \leq 0.7$) spectra extending into the submillimetre regime
(e.g. Gear et al. 1994). There have been few investigations of the
radio--millimetre spectral indices of lobe-dominated quasar
cores. This is due largely to the fact that their 5 GHz flux-densities
are typically in the range $10-100$ mJy and, given that their core
spectra are not flat, most have been below the sensitivity limit of
many millimetre and submillimetre telescopes. In a study of six
lobe-dominated quasars, Antonucci, Barvainis \& Alloin (1990) find
that the typical spectral index between 22 GHz and 230 GHz is $\alpha
\approx 1.0$. Van Bemmel \& Bertoldi (2001) found the 5--250 GHz
spectral indices of three radio quasars to all be $\approx 1$. Athreya
et al. (1997) compare radio galaxy, lobe-dominated and core-dominated
quasar cores and predict that most lobe-dominated quasar cores are
flat $\alpha \approx 0$ at GHz frequencies, but steepen at $\sim 30$
GHz (rest-frame). In Sec.\ 2 we showed that the core spectra of
3C~268.4 and 3C~298 are also likely to have $\alpha \approx
1$. Another one of our quasar sample, 3C~280.1, has an IRAM 1.25mm
flux-density of $3.5 \pm 1.7$ (Andreani et al. 2001). This suggests
its core spectral index from 5 GHz to 1.25 mm is $\alpha \approx 0.7$,
again indicating that the $850 \umu$m emission must contain a significant contribution from dust. The one example of a flat ($\alpha \approx 0.3$)
millimetre spectral index for a lobe-dominated radio source is the
unusual radio galaxy 6C~0902+34 (as discussed by A01). The radio
properties of this object suggest its jet axis is at a small angle to
our line-of-sight and dust obscuration prevents us from observing the
nuclear source and classifying it as a quasar (Carilli 1995).
Therefore, the available evidence suggests that the core millimetre
spectral indices of (lobe-dominated) radio quasars are mostly steep
($\approx 1$), but with a significant scatter such that some sources
will have flatter core spectra.

\begin{figure}
%\hspace{-0.7cm}
\epsfxsize=0.45\textwidth \epsfbox{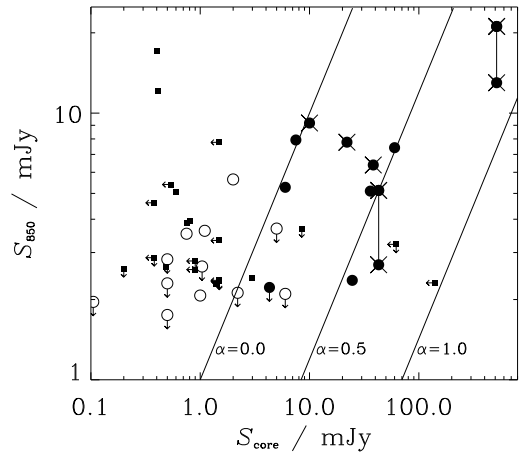}
{\caption[junk]{\label{fig:core} 850 $\umu$m flux-density $S_{850}$
against radio core flux-density at 5 GHz $S_{\rm core}$ for radio
quasars and radio galaxies (symbols as in Fig. 2). The curves
correspond to the predicted contribution of core synchrotron radiation
to $S_{850}$ as a function of $S_{\rm core}$ and core spectral index
$\alpha$.}}
\end{figure}

In Fig. \ref{fig:core} we plot $S_{850}$ against the radio core
flux-densities at 5 GHz given in Table 1. As expected, the radio
quasars have cores which are brighter than those of the radio galaxies
by a factor $\sim 10$. In Fig. \ref{fig:core} we also plot lines
corresponding to the expected sub-mm flux-density of the core as a
function of core flux-density for a range of spectral indices between
5 GHz and 850 $\umu$m. 3C~298 has a core flux a factor of ten greater
than any of the other quasars in our study. As we saw in the previous
section, its core almost certainly makes a significant contribution to
its 850 $\umu$m flux of $\approx 10$ mJy. If the five quasars on the
$\alpha=0.5$ line all have spectra with $\alpha \approx 1$ then the
synchrotron contribution to their 850 $\umu$m fluxes is $\sim 1$ mJy,
well below their detected fluxes with the exception of 3C 205. Flatter
spectral indices of $\alpha \approx 0.5$ would be necessary for the
cores to be responsible for the detections of all these quasars. In
reality, it is likely that there is a spread in the GHz to millimetre
spectral indices of the quasar cores such that a few of these quasar
submillimetre detections might well be influenced by synchrotron
emission.

Five of the radio quasars (3C~191, 3C~268.4, 3C~298, 3C~318 and
6C~1045+3513) are detected at shorter far-infrared wavelengths and
from their SEDs we are confident that they are dusty hyperluminous
objects (Sec.\ 2). For the other quasars, none of the 450 $\umu$m
non-detections are sensitive enough for us to be able to exclude the
presence of a thermal dust spectrum being responsible for the observed
850 $\umu$m emission. 3C~181 and 3C~432 have very weak cores, which
are steep ($\alpha>0.5$) at GHz frequencies and therefore their cores
cannot be significant at 850 $\umu$m, so we consider these as further
secure detections of dust. The IRAM 1.25mm flux of 3C~280.1 suggests
that its core component cannot account for all the observed 850
$\umu$m flux. The remaining two submillimetre bright quasars (3C~205
and 3C~270.1) lie close to the $\alpha=0.5$ line in
Fig. \ref{fig:core} and we cannot yet be certain that we have detected
dust in these objects. Given the available evidence of the likely
range of core spectra, we suspect that one or two of these
submillimetre detections will be dominated by synchrotron
emission. Sensitive millimetre observations of these sources are
needed to differentiate between synchrotron and dust. We thus repeated
the statistical tests performed in Section 2, now assuming,
conservatively, that the submillimetre detections of both of these
quasars are due solely to synchrotron emission: the detections were
replaced with non-detections of $0.0 \pm 1.0$ mJy. The mean quasar
$S_{850}$ is now $5.2 \pm 1.3$, still a factor of 4 greater than the
mean radio galaxy $S_{850}$ of $1.2 \pm 0.6$ The survival analysis
tests confirm that a statistical difference between the submillimetre
fluxes of radio galaxies and quasars remains at 95\% significance.

\section{Why do some radio quasars have high submillimetre luminosities?}

We have shown that core synchrotron emission may make
a small contribution to the $850 \umu$m fluxes of 3C radio quasars, but
cannot explain all the submillimetre difference between quasars and
radio galaxies. In Fig. \ref{fig:l151} we showed that there is no
systematic difference in the radio luminosities of the two types of
object. In this section we consider whether there are any other
differences between radio galaxies and quasars which may help to
explain the difference in their submillimetre luminosities.
The key to following our arguments is that we feel that
in any realistic unified scheme we must go beyond the
simplest model in which the probability
of viewing a quasar nucleus is unity for a viewing angle
$\theta \leq \theta_{\rm trans}$, with $\theta_{\rm trans}$ constant, and
zero at $\theta > \theta_{\rm trans}$.

\subsection{The effect of the intrinsic luminosity of the quasar}

As discussed in Sec.\ 1 the idea that radio galaxies and radio quasars
with similar low-frequency radio luminosities should have similar
distributions in a second orientation-independent quantity is not
quite true. For example, Simpson (1998) has quantified the factor by
which quasars should have higher [O\,{\small III}] line strengths than
radio galaxies, matched in radio luminosity, in the context of the
receding-torus model for AGN (Lawrence 1991).  Considering a
population of quasars with a spread in optical luminosity, those
quasars with higher optical luminosities have a larger torus opening
angle due to the radius at which dust is sublimated being
larger. Taking into account the scatter in the correlation between
radio and optical luminosity, this means that sources with higher
optical luminosities can be viewed within a larger opening angle and
are thus more likely to be classified as quasars. An additional effect
is that the larger opening angles allow a greater fraction of the
ionizing continuum to escape, however this is not the dominant effect
in this case. Simpson (1998) showed that this leads to a sample of
radio quasars having a mean optical luminosity, as probed by their
[O\,{\small III}] line luminosity, approximately twice that of a
sample of radio galaxies of the same radio luminosity.

\begin{figure}
%\hspace{-0.7cm}
\epsfxsize=0.45\textwidth
\epsfbox{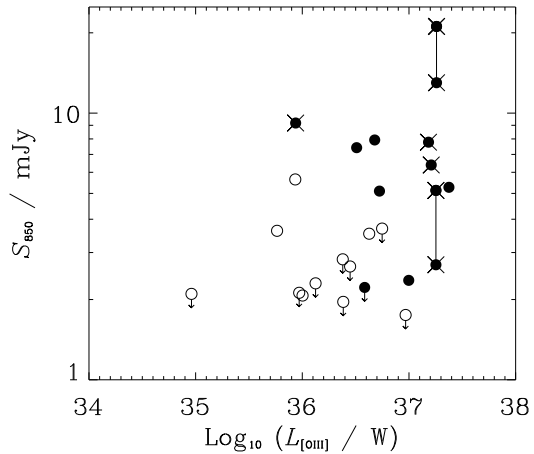}
{\caption[junk]{\label{fig:lline} 850 $\umu$m flux-density $S_{850}$
against narrow [O\,{\small III}] emission line luminosity $L_{\rm
[O\,{\small III}]}$ for radio quasars and radio galaxies (symbols as
in Fig. 2).}}
\end{figure}

In Fig. \ref{fig:lline} we plot $S_{850}$ against the [O\,{\small
III}] emission line luminosity $L_{\rm{[O\,{\small III}]}}$, which is
emitted isotropically, believed to be produced outside any obscuring
torus, and which is a good indicator of the ionizing luminosity
(Simpson 1998). Correlation tests show that $S_{850}$ and
$L_{\rm{[O\,{\small III}]}}$ for the radio quasars and radio galaxies
are positively correlated at 95\% significance. It is also apparent
that the quasars do typically have higher $L_{\rm{[O\,{\small III}]}}$
than the radio galaxies. For the quasars the mean $\log
(L_{\rm{[O\,{\small III}]}} / {\rm W)}$ is $36.9 \pm 0.1$ and for the
radio galaxies it is $36.2 \pm 0.2$. Thus there is a factor of five
difference in the mean [O\,III] luminosities of the two types of
object in our sample. A similar difference between $L_{\rm{[O\,{\small
III}]}}$ for the highest radio luminosity quasars and radio galaxies
is also evident from Fig.\ 5 of Jackson \& Rawlings (1997). In that
study, all the objects have [O\,{\small III}] line observations made
with the same wide slit, thereby eliminating uncertainties due to
inhomogeneous datasets. This difference in emission line luminosity
(and consequently ionizing luminosity) could explain the difference of
a factor a few in the average luminosities of quasars and radio
galaxies in the mid- and far-infrared in studies with IRAS and ISO
(Heckman et al. 1994; Hes et al. 1995; van Bemmel, Barthel \& de
Graauw 2000; Meisenheimer et al. 2001), since this radiation is
certainly dominated by nuclear emission reprocessed by the dust torus.

Could such an effect also be responsible for the similar factor
difference in the submillimetre luminosities of radio quasars and
radio galaxies? This would require a tight correlation between the
submillimetre and optical continuum emission from quasars. The
analysis of Simpson (1998), predicting a factor of two difference in
the [O\,{\small III}] luminosities of radio quasars and radio
galaxies, used the known scatter of 0.6 dex in the radio--optical
correlation (Serjeant et al. 1998; Willott 2000). Therefore, we would
expect the scatter exhibited in the radio--submillimetre correlation
to be of the same order if the submillimetre and optical continua are
tightly correlated. A power law fit to the marginal correlation
between $L_{151}$ and $S_{850}$ in Fig.2 gives a linear slope with a
scatter of 0.3 dex. The scatter may be slightly larger than this
because of the large number (40\%) of non-detections (in the above
analysis non-detections were set to their 1$\sigma$ upper
limits). However, within the limited range in $L_{151}$ available in
our study, we find the scatter in the radio--submillimetre correlation
to be no larger than that of the radio--optical correlation.

A tight correlation between submillimetre and optical luminosities
could probably only come about by direct heating of the dust
responsible for the submillimetre emission by the AGN (as is known to
be the case for the mid-infrared emission, and has been discussed
previously by Willott et al.\ 2000b). In this case the scatter could
be as small as that seen in the correlation between optical and
mid-infrared luminosities of optically-bright, low-redshift quasars,
i.e. 0.2 dex from the study of Rowan-Robinson (1995).  If the
far-infrared emission is due to dust heated by an intense starburst
rather than by the quasar, a larger scatter is predicted; indeed the
larger scatter, of 0.4 dex, in the far-infrared versus optical
luminosity plot from Rowan-Robinson (1995) was used as an argument
that: (i) the far-infrared emission from the low-redshift quasars is
dominated by starburst-heated dust; and (ii) the scaling relation
between optical and far-infrared luminosities arises from both
quantities being correlated with a third variable, i.e.\ the black
hole and/or galaxy mass. However, in the quasar-heating hypothesis,
the far-infrared luminosity is strongly dependent upon the extended
dust distribution (e.g. Granato \& Danese 1994) and would indeed be
expected to show a larger scatter than the mid-infrared--optical
ratio. It may seem surprising that objects with such large quantities
of dust have such a small scatter in these ratios because of the
effects of dust-reddening of the optical emission. However, in both
the Rowan-Robinson sample and the radio-loud sample discussed in this
paper, objects would not be classified as quasars if they were heavily
reddened ($A_V \gtsimeq 3$).

Finally in this section, we note that despite the quantitative
agreement between the predictions of the receding-torus model and the
submillimetre difference between radio quasars and radio galaxies,
other physical effects seem likely to be important.  Note, for
example, that the high submillimetre luminosity of 6C~1045+3513 is an
anomaly in the receding-torus model, since it has a value of
$L_{\rm{[O\,{\small III}]}}$ below that of all the 3C radio galaxies.

\subsection{The effect of the radio source age}
\label{age}

\begin{figure}
\epsfxsize=0.45\textwidth
\epsfbox{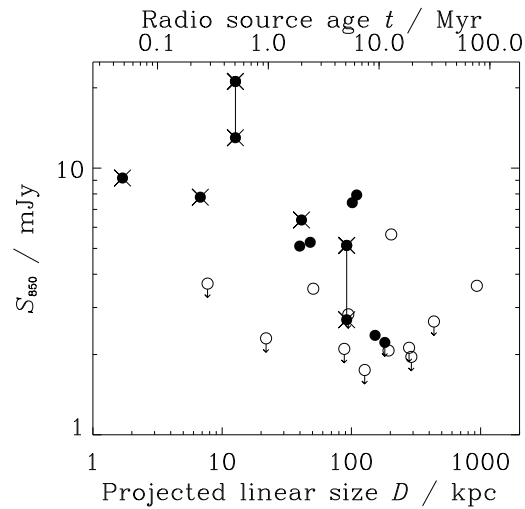}
{\caption[junk]{\label{fig:d} 850 $\umu$m flux-density $S_{850}$
against radio source projected linear size $D$ for radio quasars and
radio galaxies (symbols as in Fig. 2). There is a significant
anti-correlation present for the quasars. The upper axis shows how the
ages of radio sources approximately relate to $D$, according to the
model of Willott et al.  (1999) (see Sec.\ 4.2 for details).}}
\end{figure}

Fig. \ref{fig:d} plots $S_{850}$ against the projected linear sizes
$D$ of the radio sources. According to unified schemes, quasars will
typically have projected linear sizes a factor of $\approx 1.6$
smaller than those of radio galaxies, due to the smaller angles
between their jet axes and our line-of-sight.  (assuming $\theta_{\rm
trans}=53^{\circ}$ from Willott et al. 2000a). Taking account of this
effect, the distributions in $D$ for the quasars and radio galaxies
are fairly similar. However, a striking feature of this plot is that
the quasars appear to show an anti-correlation between $S_{850}$ and
$D$. The correlation tests performed (Kendal's Tau and Cox's) show
these quantities to be anti-correlated with significances of 95 and
99\%, respectively. Including the radio galaxies as well, and
accounting for projection effects, the significance of this
correlation is only marginal (92\% and 50\%), largely because of the
introduction of a large number of submillimetre upper limits.

An important point to consider here is whether this anti-correlation
could be due to contamination of the sub-millimetre emission by
synchrotron from the lobes or hotspots. This is because the SCUBA beam
at 850 $\umu$m has a 13 arcsec FWHM, corresponding to a physical size
at these redshifts of 110 kpc. Hence sources with $D$ less than this
value will have their hotspots within the SCUBA beam. However, one
should note that the five radio quasars with detections at shorter
wavelengths indicated by the crosses are also mostly the smallest
quasars. In these objects we have evidence for a thermal dust spectrum
and high far-IR luminosities. It can also be seen that five radio
galaxies have $D<100$ kpc and these have significantly smaller
sub-millimetre fluxes than the quasars. In the orientation-based
unified scheme there is no reason why quasars should have brighter
extended synchrotron emission than radio galaxies. Therefore, although
we cannot rule out some lobe/hotspot contamination in the smallest
sources it cannot be the driving force behind the observed
correlation.

3C~298, 3C~318, 6C~1045+3513, and more marginally 3C~191, are all
compact steep-spectrum (CSS) radio sources, which are believed to be
small due to their jets being triggered relatively recently ($ \sim
10^6$ years; Fanti et al. 1995). Previous studies of the far-infrared
properties of CSS sources have shown them to be indistinguishable from
larger, presumably older, radio sources (Heckman et al. 1994; Fanti et
al. 2000). However our study is quite different from these which were
limited to low redshifts (and hence luminosities) due to the
sensitivity limitations of IRAS and ISO, and it is plausible that the
differences only occur at higher luminosities, higher redshifts and/or
longer wavelengths (e.g.\ because the emission is dominated by
starburst-heated cold-dust emission rather than AGN-heated hot-dust
torus emission).

It is also interesting to note that 3C~298, 3C~318 and 6C~1045+3513
are reddened quasars, indicating significant quantities of dust in
their host galaxies. Also, 3C~191, 3C~268.4 and 3C~298, show very
strong CIV associated absorption and 6C~1045+3513 has a CIV broad
absorption system. Baker et al. (2001) show that there is an
anti-correlation between the strength of associated absorption systems
and radio linear size and also that the redder quasars tend to have
the strongest line absorption. It has also been noticed that, amongst
radio-loud quasars, broad absorption lines are more commonly observed
in the smallest radio sources (Becker et al. 2000). These observations
indicate an evolutionary picture where the absorbing material is
cleared as the radio sources get older (Baker et al. 2001). This is
similar to the evolutionary scheme of Sanders et al. (1988) where
quasars begin their lives in a dust-enshrouded phase.

The projected linear size $D$ of a radio quasar or radio galaxy can be
related approximately to the time since the jets were triggered $t$,
since the relationship between these variables is only weakly
dependent upon the radio luminosity, and the objects under
consideration span only 1.5 dex in this quantity. Using the simple
model of radio source evolution in Willott et al. (1999) we can, at
least approximately, transform the linear size axis in Fig.
\ref{fig:d} to the time since the quasars were triggered (the radio
galaxies have ages on average a factor of 1.6 smaller at a given $D$
than shown here, due to projection effects).  Fig. \ref{fig:d}, and
the anti-correlation it exhibits, shows that the most
submillimetre-luminous quasars had their radio sources triggered
$<10^{7}$ years ago.  This means that the processes controlling the
submillimetre/far-infrared luminosity $L_{\rm FIR}$ are synchronized
with the jet-triggering event and that $L_{\rm FIR}$ declines on a
$\sim 10^{7}$ yr timescale.

Although high values of $L_{\rm FIR}$ for an AGN are generally
interpreted as evidence for high rates of star formation, it is
possible that, as explored in Sec.\ 4.1 (see also Willott et
al. 2000b; Rowan-Robinson 2000), even the cool dust is directly heated
by the AGN. Any evidence for star formation is indirect, being based
on the large gas masses observed in CO in a handful of cases, but
mostly deduced from the large dust masses required by any
submillimetre-detected thermal source (e.g.\ Hughes, Dunlop, \&
Rawlings 1997).  With a standard dusty disc model (e.g. Efstathiou \&
Rowan-Robinson 1995) there are difficulties in heating the cool dust
at large distances from the nucleus, due to the fact the disc is
optically thick and shields the distant dust from the ionizing
radiation. A more efficient way of heating large amounts of cool dust
by an AGN, come from models where the dust has a more isotropic
distribution with a very high covering factor (e.g. Granato, Danese \&
Franceschini 1996). This may occur preferentially in
recently-triggered AGN if the dusty gas fueling the nucleus is not in
a regular torus distribution. In addition, models for shocks caused by
advancing radio components suggest that dust grains cannot survive for
much longer than $\sim 10^{7} \rm yr$ in such environments (De Young
1998).  Jarvis et al.\ (2001b) summarise other evidence that expanding
radio sources modify the physical conditions of their gaseous
environments as the radio-emitting components expand outward after jet
triggering. There are therefore two reasons why in a scenario in which
quasar-heated dust dominates, one expects $L_{\rm FIR}$ to decline on
a $\sim 10^{7} \rm yr$ timescale. Note that the fact that the
brightest submillimetre sources are reddened quasars with line
absorption is consistent with this picture, since with this dust
distribution there is a high probability of our line-of-sight to the
nucleus intercepting a dust cloud. It is of course possible that the
dust distribution could be influenced by previous episodes of
star-formation.

It is also possible that the submillimetre emission in the
hyperluminous sources is dominated by starburst-heated dust, in which
case the implication is that the jet-triggering process is accurately
synchronized with an intense burst of star formation.  The timescale
for such star formation events would have to be $\ltsimeq 10^7$ years
for the far-infrared (300 $\umu$m) luminosity to peak $\approx 10^7$ yr
after the onset of the starburst (Efstathiou, Rowan-Robinson \&
Siebenmorgen 2000). Line ratios from mid-infrared spectroscopy of
nearby starburst galaxies suggest that the duration of the bursts in
such galaxies are $\sim 10^7$ yr (Thornley et al. 2000) and for
ultraluminous infrared galaxies $\sim 2 \times 10^7$ yr (Genzel et
al. 1998). Thus the starburst timescales are quite consistent with the
hypothesis that a starburst and a jet-producing AGN were triggered
synchronously, and the anti-correlation seen in Fig. \ref{fig:d}.

A starburst of such a brief duration, forming stars at $\sim 1000
~{\rm M_{\sun}}$ yr$^{-1}$ as implied by the submillimetre luminosity,
would form $\sim 10^{10}~ {\rm M_{\sun}}$ of stars over its
lifetime. The total stellar mass of a powerful radio galaxy with a
luminosity of $5{\rm L_{\star}}$ (Jarvis et al. 2001a) is $5 \times 10^{11}
{\rm M_{\sun}}$. Therefore such a star formation episode would form only a
few per cent of the total stellar mass in a massive elliptical galaxy
and therefore this phase is not a major event in the star-formation
histories of the quasar host galaxy. This is consistent with other
evidence, such as the small scatter in the $K-z$ relation (Jarvis et
al. 2001a) and the existence of very red radio galaxies at $1<z<2$
(Dunlop et al.\ 1996; Dunlop 1999; Willott, Rawlings \& Blundell
2001), that the bulk of the stars in powerful radio galaxies were
formed at higher redshifts of $z>3$.

Regardless of whether the dust is heated by a quasar or a starburst or
a mixture of the two, it seems that Fig. \ref{fig:d} provides some
support for the Blundell \& Rawlings (1999) conjecture that radio
source youth may be an important factor in explaining the high
submillimetre luminosities of distant radio galaxies. An important
question is how this finding relates to the properties of the more
distant ($z > 2$) systems studied by A01.  We again defer a more
detailed discussion to a future paper (Rawlings et al.\, in prep.) but
note here that the fact that the submillimetre luminosities of the
$z<2$ radio quasars are similar to the radio galaxies at $z>2$ is not,
by itself, a strong argument that their host galaxies are in similar
evolutionary states. It is certainly plausible that the
higher-redshift objects are forming significant fractions of the
stellar mass of a giant elliptical galaxy, requiring simply that the
duration of the starburst is $ \gg 10^7$ yr. Direct measurements of
gas mass, e.g.\ by CO observations, would be very important for
determining the evolutionary states and the main heating source of the
cool dust for the $z \sim 1.5$ objects studied here. It is hard with
submillimetre continuum measurements alone to determine the
relationship between these objects and the high-redshift radio
galaxies studied by A01. The discovery of huge reservoirs of molecular
gas in some very high-redshift radio galaxies (Papadopoulos et
al. 2000; Alloin, Barvainis \& Guilloteau 2000) supports a
starburst-heating model for the bulk of the submillimetre continuum in
these objects, and the lack of such reservoirs at $z \sim 1.5$, if
observed, would suggest that quasar-heating might become more
prevalent at lower redshifts.

\section{Concluding remarks}

We have performed a submillimetre survey of radio quasars in the
redshift range $1.37<z<2$ and detect ten out of eleven quasars,
implying submillimetre/far-infrared luminosities $\gtsimeq 4 \times
10^{12} {\rm L_{\sun}}$. We argue that only one or two of these
detections are dominated by core synchrotron emission. For five
quasars, detections at shorter wavelengths confirm the presence of a
thermal continuum and a large mass ($\sim 10^8 {\rm M}_{\sun}$) of
cool dust.

We find the radio quasars have higher submillimetre luminosities by a
factor of $\sim 5$ than radio galaxies of the same radio luminosity
and redshift, a factor which cannot be reduced below $\sim 2$ by
accounting for possible synchrotron contamination. At $z \sim 1.5$
radio quasars are intrinsically brighter than radio galaxies at
submillimetre wavelengths.  This argues against the simplest unified
scheme for radio sources in which the probability of viewing a quasar
nucleus is unity for a viewing angle $\theta \leq \theta_{\rm trans}$,
with $\theta_{\rm trans}$ constant, and zero at $\theta > \theta_{\rm
trans}$. It is, however, in quantitative agreement with a
receding-torus model in which $\theta_{\rm trans}$ depends on quasar
optical luminosity, but only if there is a close relationship between
optical luminosity and $L_{\rm FIR}$. The implication is that
quasar-heated dust dominates $L_{\rm FIR}$ for powerful radio quasars
at $z \sim 1.5$.

There is a significant anti-correlation between $L_{\rm FIR}$ and
radio source age in the sense that hyperluminous quasars tend to be
associated with young ($<10^7$ yr) sources at $z \sim 1.5$.  This
means that the processes controlling $L_{\rm FIR}$ are synchronized
with the jet-triggering event and that $L_{\rm FIR}$ declines on a
$\sim 10^{7}$ yr timescale. There is evidence for a small amount of
obscuration in the most submillimetre-luminous quasars from reddening
(3C~298, 3C~318 and 6C~1045+3513) and/or prominent associated
absorption (3C~191, 3C~268.4, 3C~298) or broad absorption lines
(6C~1045+3513). We speculate that shortly after jet triggering, dust
is more widely distributed than at later times. This distributed dust
could be heated by the quasar nucleus and/or a short-lived
synchronized starburst and may be destroyed by shocks associated with
the expanding radio source. Any such starburst would form only a few
per cent of the total stellar mass in agreement with the evidence for
dominant old stellar populations in $z \sim 1.5$ radio galaxies.

\section*{Acknowledgements}

We thank the anonymous referee for some very useful suggestions. We
are very grateful to the staff at the Joint Astronomy Centre for their
help with the observations. The JCMT is operated by the Joint
Astronomy Centre on behalf of the U.K. Particle Physics and Astronomy
Research Council. The William Herschel Telescope is operated on the
island of La Palma by the Isaac Newton Group in the Spanish
Observatorio del Roque de los Muchachos of the Instituto de
Astrofisica de Canarias. This research has made use of the NASA/IPAC
Extra-galactic Database, which is operated by the Jet Propulsion
Laboratory, Caltech, under contract with the National Aeronautics and
Space Administration. CJW thanks PPARC for support.

\appendix

\section{6C~1045+3513 - A submillimetre-luminous reddened BAL quasar}

In this appendix we present a new optical spectrum of 6C~1045+3513,
which is one of the brightest radio quasars at submillimetre
wavelengths found in this paper ($S_{850}=9.18 \pm 1.10$ mJy). It is
also detected at the $3 \sigma$ level with the short wavelength array
($S_{450}=21.3 \pm 6.8$ mJy).

6C~1045+3513 was observed with the ISIS double-beam spectrograph on
the William Herschel Telescope on April 16 2001. Conditions were
photometric with 1.0 arcsec seeing and a 1.0 arcsec slit was used
giving a dispersion resolution of 8 \AA. The telescope was pointed at
the position of the compact radio source (10:48:34.25 +34:57:25.0 -
J2000.0). The total integration time of the observation was 1200
seconds. The data were reduced in the IRAF package using the same
method as in Willott et al. (1998). The red and blue arm spectra were
joined together, averaging over 35 \AA~ where the signal-to-noise
ratio of the spectra were approximately equal (6200 \AA).

\begin{figure}
%\vspace{0.5cm}
\hspace{-0.7cm}
\epsfxsize=0.5\textwidth 
\epsfbox{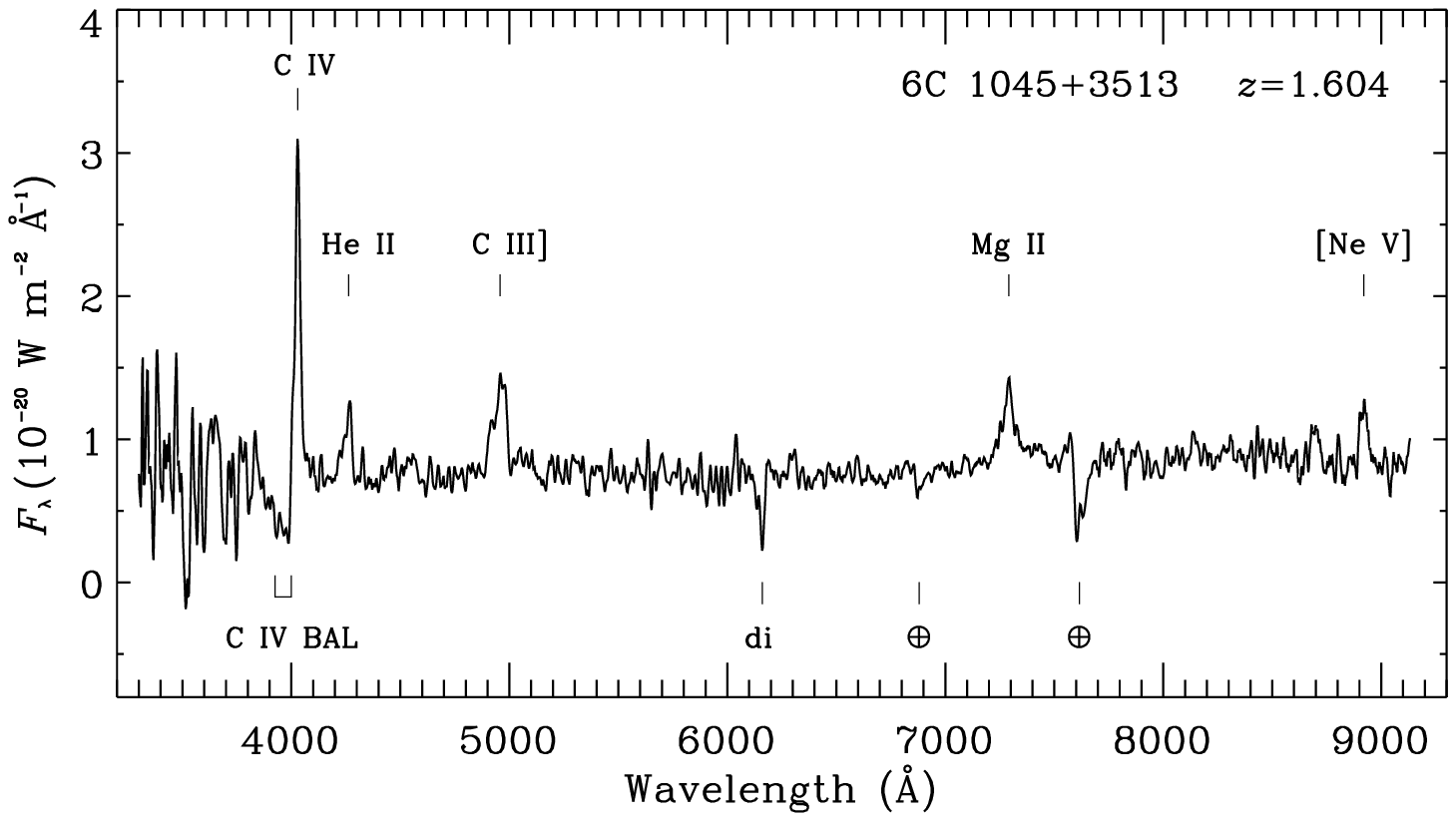}
\vspace{0.5cm}
{\caption[junk]{\label{fig:spec} Optical spectrum of the reddened
quasar 6C~1045+3513 with emission lines labelled. The CIV broad
absorption feature is also marked. The absorption features marked with
$\oplus$ are due to atmospheric absorption and that marked `di' is an
artefact caused by the dichroic.}}  
\end{figure}

The target is clearly identified in the two dimensional spectrum and
appears to be spatially unresolved. In Figure \ref{fig:spec} we show
the optical spectrum extracted in a 1 arcsec aperture. The object has
a red continuum and several strong emission lines (see Table A1). The
red continuum has a spectral index of 2.3 and is well fit by a quasar
spectrum reddened by $A_{\rm V}=1.5$ (assuming a Milky-Way extinction
curve). The positions of the emission lines are consistent with a
redshift for 6C~1045+3513 of $z=1.604$, extremely close to the
tentative value reported by Rawlings et al. (2001). The C\,{\small IV}
and He\,{\small II} emission lines are both narrow (FWHM $\approx
2000~ {\rm km~s}^{-1}$).  The Mg\,{\small II} and C\,{\small III}]
emission lines appear to have narrow central profiles with broad
wings, which together with the weak or absent broad wings in
C\,{\small IV} and He\,{\small II}, are consistent with broad-line
region reddening of the same amount as the continuum fitting suggests.

There is very high equivalent width C\,{\small IV} absorption on the
blue wing of the emission line. This feature is very broad with a
width of at least $6000~ {\rm km~s}^{-1}$. There is no corresponding
broad absorption of Mg\,{\small II}, so 6C~1045+3513 is classified as a
high-ionization broad absorption line (BAL) quasar. Unfortunately, our
data are not sensitive enough or with sufficient spectral resolution
to determine the structure of the absorption trough. 6C~1045+3513 is
one of the most radio-luminous quasars to show the BAL phenomenon,
with a radio luminosity similar to that of the BAL QSO FIRST
J101614.3+520916 (Gregg et al. 2000) .

6C~1045+3513 is thus remarkably similar to 3C~318 (Willott et
al. 2000b). Both objects have a small degree of reddening which
suppresses the rest-frame UV broad line strengths to below those of the
narrow lines. Both are also very luminous submillimetre sources (with
detections at both 850 $\umu$m and 450 $\umu$m) and both are
compact ($D<10$ kpc) radio sources with similar estimated source ages.
A possible reason for these similarities is discussed in Sec.\ 4.2.

\begin{table}
\footnotesize
\begin{center}
\begin{tabular}{lcccc}
\hline
\mc{1}{l}{Line} &\mc{1}{c}{$\lambda_{\mathrm obs}$} &\mc{1}{c}{$z_{\mathrm em}$} &\mc{1}{c}{FWHM}
&\mc{1}{c}{Flux/$10^{-19}$ }  \\
\mc{1}{c}{ } &\mc{1}{c}{(\AA)} &\mc{1}{c}{ } &\mc{1}{c}{(km s$^{-1}$)} &\mc{1}{c}{(W m$^{-2}$)} \\

\hline
C\,{\footnotesize IV}          $\lambda 1549$ &  4029 $\pm$ 4 & 1.601 & 2000  &    18 $\pm$ 4   \\
He\,{\footnotesize II}         $\lambda 1640$ &  4263 $\pm$ 6 & 1.599 & 2400  &\f   3.6 $\pm$ 1.3 \\
C\,{\footnotesize III}]        $\lambda 1909$ &  4906 $\pm$ 6 & 1.597 & 4300  &\f   8.2 $\pm$ 1.0 \\
Mg\,{\footnotesize II}         $\lambda 2799$ &  7291 $\pm$ 2 & 1.605 & 2900  &\f   9   $\pm$ 4   \\
$\rm{[\mbox{Ne\,{\footnotesize V}}]}~  \lambda 3426$ &  8920 $\pm$ 3 & 1.604 & 1300  &\f   4.6 $\pm$ 1.6 \\
\hline
\end{tabular}

{\caption[junk]{\label{tab:lines} Emission line data for 6C~1045+3513
from our optical spectrum. Fluxes have been aperture corrected.}}

\normalsize
\end{center}
\end{table}

%EW data
%&  80 $\pm$ 20
%&  11 $\pm$ 4    \f
%&  23 $\pm$ 4    \f
%&  23 $\pm$ 11
%&  13  $\pm$ 5   \f

\end{document}